\documentclass[onecollarge,natbib]{svjour2}
\bibpunct{[}{]}{,}{n}{}{,} 
\smartqed  
\usepackage{graphicx}
\journalname{Few-Body Systems (APFB2011)}
\begin{document}

\title{\boldmath
Parity-violating asymmetry in 
$\gamma d \to \vec{n}p$ with a pionless effective theory
}


\author{J. W. Shin      \and
        S.-I. Ando      \and
	C. H. Hyun      \and
	S. W. Hong
}


\institute{J. W. Shin \at
              Department of Physics, Sungkyunkwan University,
              Suwon 440-746, Korea \\
              \email{shine8199@hanmail.net}           
           \and
           S.-I. Ando \at
              Department of Physics Education,
              Daegu University, Gyeongsan 712-714, Korea \\
	      \email{sando@daegu.ac.kr}
           \and
           C. H. Hyun \at
              Department of Physics Education,
              Daegu University, Gyeongsan 712-714, Korea \\
	      \email{hch@daegu.ac.kr}
	   \and
           S. W. Hong \at
              Department of Physics and Department of Energy Science, 
	      Sungkyunkwan University, Suwon 440-746, Korea \\
	      \email{swhong@skku.ac.kr}
}

\date{Received: date / Accepted: date}

\maketitle

\begin{abstract}
Nuclear parity violation is studied with polarized neutrons
in the photodisintegration of the deuteron at low energies.
A pionless effective field theory with di-baryon fields 
is used for the investigation.
Hadronic weak interactions are treated by parity-violating
di-baryon-nucleon-nucleon vertices, which have undetermined
coupling contants.
A parity-violating asymmetry in the process is calculated
for the incident photon energy up to 30~MeV.
If experimental data for the parity-violating asymmetry become
available in the future, we will be able to
determine the unknown coupling contants in the parity-violating
vertices.

\end{abstract}

\section{Introduction}
\label{intro}

Parity-violating (PV) nuclear force has been treated 
in the framework of one-meson-exchange model, 
known as the DDH potential~\cite{DDH}.
However, recently the PV nuclear force was derived in an effective field theory (EFT)
of QCD up to one-loop order by considering two-pion-exchange 
contributions~\cite{zmhrmvk-npa05,had-plb07}.
Meanwhile, another version of EFT, a pionless EFT 
in which pions are regarded as a heavy degree of freedom 
and thus integrated out, has been demonstrated efficient for
describing both parity-conserving (PC) 
(see, {\it e.g.} \cite{ando06b,ando08}) and 
PV nuclear few-body processes at very low energies.

A PV two-nucleon process with a pionless theory 
was first studied by Savage~\cite{savage01}.
Subsequently we have calculated PV observables 
in $np \to d\gamma $ at threshold in the pionless 
EFT~\cite{hyun09,ando10,shin10}.
In our previous works, we obtained a PV Lagrangian employing
di-baryon fields which represent two-nucleon $^1S_0$ and 
$^3S_1$ states. (A similar Lagrangian was obtained independently
in \cite{pss-npa09}.) 
The weak $NN$ interaction is described in terms of 
the PV di-baryon-nucleon-nucleon ($dNN$) vertices, 
in which the two-nucleon S-wave states represented by the di-baryon fields 
make a transtion to the P-wave states represented by the two nucleon fields, 
whereas the interactions are assumed to have unknown weak $dNN$ coupling constants. 

Recently we considered the neutron spin polarization $P_{y'}$ 
along the direction $\hat{y}'$ 
\footnote{Conventions for the directions are illustrated in \cite{rus60}.}
in $\gamma d \to \vec{n} p$ with the pionless EFT \cite{ando11},
where $\vec{n}$ refers to a polarized neutron.
The neutron spin polarizations along the other directions 
$\hat{x}'$ and $\hat{z}'$ 
vanish with only the PC interactions,
but they can be nonvanishing with the PV interactions.
Interference terms between PC and PV amplitudes can give 
non-zero contributions to the neutron spin asymmetries.

Here, we report on our calulation of the PV 
neutron spin asymmetry in $\gamma d \to \vec{n}p$ process
in terms of the undetermined PV coupling constants
in the pionless EFT.
We investigate the energy dependence of the asymmetry $P_{z'}$
at a few colatitude angles with the photon energies 
up to 30 MeV.
If measurements of the PV asymmetry can be made
in future,
we will be able to fix the undetermined PV coupling constants
in the theory.

\section{PV Lagrangian}
\label{sec:1}

A coefficient in an effective Lagrangian, 
known as a low-energy constant (LEC), 
is assumed to include contributions 
from degrees of freedom that are integrated 
out of a theory.
In the pionless theory, because even the pion is 
integrated out, an interaction is described only by
the nucleon-nucleon contact terms with undetermined LECs.
In the case of pionless theory with di-baryon fields, 
we assume that a PV $dNN$ vertex subsumes the PV $NN$ interactions 
of the integrated-out degrees of freedom.
Introducing dimensionless PV LECs 
$h^{\Delta I}_d$,
we have a PV $dNN$ Lagrangian for the $\Delta I=0$ 
part as~\cite{shin10}
\begin{eqnarray}
{\cal L}^0_{\mbox{\tiny PV}} &=&
\frac{h^{0 s}_d}{2 \sqrt{2\, \rho_d\, r_0}\, m_N^{5/2}} 
s^\dagger_a\, N^T 
\sigma_2 \sigma_i  \tau_2 \tau_a 
\frac{i}{2} \left(\stackrel{\leftarrow}\nabla - 
\stackrel{\rightarrow}\nabla \right)_i N +{\rm h.c.} \\
& & + 
\frac{h^{0 t}_d}{2 \sqrt{2} \rho_d\, m_N^{5/2}} \,
t^\dagger_i\, N^T \sigma_2 \tau_2 
\frac{i}{2} \left(\stackrel{\leftarrow}\nabla - 
\stackrel{\rightarrow}\nabla \right)_i N
+{\rm h.c.}\,,
\end{eqnarray}
where $s_a$ ($r_0$) and 
$t_i$ ($\rho_d$) are di-baryon fields (effective ranges) 
in the $^1S_0$ and $^3S_1$ channel, respectively.
Spin-isospin operator $\sigma_2 \sigma_i \tau_2 \tau_a$ 
with the operator 
($i\nabla$)
in Eq. (1) projects a two-nucleon system to $^3 P_0$ state. 
The PV vertex given in Eq. (1) therefore generates a
$^3 P_0$ admixture in the $^1 S_0$ state. 
Similarly, $\sigma_2 \tau_2$ 
with the operator ($i\nabla$) in Eq. (2) is the
projection operator for $^1 P_1$ state, and thus the Lagrangian
mixes $^1 P_1$ state with the $^3 S_1$ state.
For the $\Delta I= 1$ part, we have $^3 P_1$ admixture to the $^3 S_1$
state, so the Lagrangian reads
\begin{eqnarray}
{\cal L}^1_{\mbox{\tiny PV}} =
i \frac{h^1_d}{2 \sqrt{2} \rho_d\, m_N^{5/2}} \,
\epsilon_{ijk}\, t^\dagger_i\, N^T \sigma_2 \sigma_j \tau_2  \tau_3
\frac{i}{2} \left(\stackrel{\leftarrow}\nabla - 
\stackrel{\rightarrow}\nabla \right)_k N
+{\rm h.c.}.
\end{eqnarray}

\section{Neutron spin polarization with PV interaction}
\label{sec:2}
\begin{figure*}
\centering
  \includegraphics[width=0.85\textwidth]{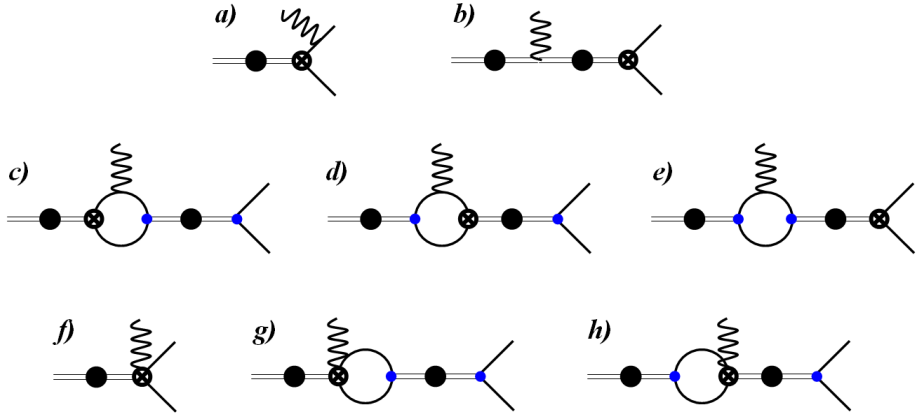}
\caption{Leading order ($Q^{0}$) PV diagrams for 
$\gamma d \to \vec{n}p$. 
Single solid line denotes a nucleon, 
a wavy line represents a photon, and a double line with a filled circle 
refers to a dressed di-baryon propagator. 
A circle with a cross represents a PV $dNN$ vertex.}
\label{fig:1}       
\end{figure*}

In Figure 1 we show the leading-order diagrams 
for the PV $\gamma d \to \vec{n}p$ process. 
From the sum of all the amplitudes, 
we can obtain the transition amplitudes involving 
the PV interactions.

For the polarization of the neutron along an axis $\hat{n}$,
we introduce the spin-isospin projection operator

\begin{eqnarray}
P_{\pm} = \frac{1}{2}(1-\tau_{3})\frac{1}{2}(1 \pm \vec{\sigma} \cdot \hat{n}).
\end{eqnarray}
Inserting the projection operator 
in the spin-isospin summation of the squared amplitude,
we obtain
\begin{eqnarray}
S^{-1} \sum_{spin}^{P} |A|^{2} 
&=& 4(|X_{MS}|^{2}+ |Y_{MV}|^{2}-2Y_{MV}ReX_{MS}) \nonumber \\
&+& 2(|X_{MV}|^{2}+ |Y_{MS}|^{2}-2Y_{MS}ReX_{MV}) \nonumber \\
&+& 3(1-(\hat{k}\cdot \hat{p})^{2})(|X_{E}|^2+|Y_{E}|^{2}-2X_{E}Y_{E}) 
\nonumber \\
&\mp& 2  \hat{n} \cdot (\hat{k} \times \hat{p})(X_{E}-Y_{E})ImX_{MV} 
\nonumber \\
&\mp& 2 (\hat{k} \cdot \hat{n}) Im f(pv1) \nonumber \\
&\mp& 2 (\hat{p} \cdot \hat{k})(\hat{k} \cdot \hat{n}) Im f(pv2) \nonumber \\
&\mp& 2 (\hat{p} \cdot \hat{n}) Im f(pv3) \nonumber \\
&\mp& 2 (\hat{p} \cdot \hat{k})(\hat{p} \cdot \hat{n}) Im f(pv4),
\label{eq:squared-amplitude}
\end{eqnarray}
where $S$ is a symmetric factor $S=2$.
Following the conventions given in \cite{rus60},
we have the incoming photons along $\hat{k} = (0,0,1)$,
relative momentum of the nucleons along 
$\hat{p} = (\sin{\theta}\cos{\phi},\sin{\theta}\sin{\phi},\cos{\theta})$,
and the neutron spin polarized along
$\hat{n} = \hat{z}' ( = \hat{p})$.
$Im f(pv1)$, $Im f(pv2)$, $Im f(pv3)$ and $Im f(pv4)$ are 
the interference terms of PC and PV amplitudes.
Because the explicit expressions of these quantities are lengthy, they will be given
in a forthcoming article \cite{sahh}.
The terms 
$X_{MV}$, $X_{MS}$, $X_{E}$, 
$Y_{MV}$, $Y_{MS}$ and $Y_{E}$ 
in Eq.~(\ref{eq:squared-amplitude}) 
are the same as those in Ref.~\cite{ando11}.
%

The neutron spin asymmetry along the $\hat{z}'$ direction vanishes 
only with the PC interactions, but, as mentioned before, 
it can be nonvanishing in the presence of the PV interactions.
We obtain the PV neutron spin asymmetry $P_{z'}$ 
for $\gamma d\to \vec{n}p$ as
\begin{eqnarray}
P_{z'} &=& 
\frac{\sigma_{+}(\theta) - \sigma_{-}(\theta)}
 {\sigma_{+}(\theta) + \sigma_{-}(\theta)} 
\nonumber \\
&=& (-2)[\cos{\theta} \{ Im f(pv1)+Im f(pv4) \} 
 +\cos^{2}{\theta}Im f(pv2)+Im f(pv3)]  
\nonumber \\
&/&  \bigg{[} 4(|X_{MS}|^{2}+ |Y_{MV}|^{2}-2Y_{MV}ReX_{MS})
+ 2(|X_{MV}|^{2}+ |Y_{MS}|^{2}-2Y_{MS}ReX_{MV}) 
\nonumber \\
& & + 3(1-\cos^{2}{\theta})(|X_{E}|^2+|Y_{E}|^{2}-2X_{E}Y_{E}) \bigg{]},
\end{eqnarray}
where $\sigma_{\pm}(\theta)$ are the differential cross sections
with the spin-isospin projection operators $P_{\pm}$.

Figure~\ref{fig:1} shows our results as functions of the photon energy
$E_\gamma^{lab}$ at $\theta=30^\circ$, 60$^\circ$, 90$^\circ$.
Because the LECs $h_d^{0t}$, $h_d^{0s}$ and $h_d^{1}$
are not known yet, we cannot determine the numerical values
of $P_{z'}$ at present. 
Instead, we define
$P_{z'} \equiv a_{0t} h^{0t}_d + a_{0s} h^{0s}_d + a_1 h^1_d$,
and Fig.~\ref{fig:1} shows $a_{0t}$, $a_{0s}$ and $a_1$ as 
functions of the photon energy.
The angle dependence of the curves for $a_{0s}$ and $a_1$
shows similar features;
as the angle 
increases, the dependence on the incoming photon 
energy becomes small.
However, the curves of $a_{0t}$ do not have angle dependence.
This is because cancellations between E1 and M1 contributions 
make the result nearly angle independent.
More detailed analysis will be reported 
in our forthcoming paper~\cite{sahh}.
Another notable point is that the magnitudes of $a_{0t}$,
$a_{0s}$ and $a_1$ are similar in the energy range considered.
In terms of DDH potential, $\rho$- and $\omega$-meson
exchange interactions correspond to the vertices represented by
the coupling constants $h^{0t}_d$ and $h^{0s}_d$, and 
$\pi$-exchange interaction corresponds to $h^1_d$.
Thus our results on $P_{z'}$ in this work seem to suggest that
the coefficients of the $\rho$- and $\omega$-meson exchange
contributions will be similar to the ones by the $\pi$-exchange 
in the DDH potential.
Therefore the calculation with DDH potential will provide a 
counter check of the results in the present work.

\section{Summary}
\label{sec:3}

In this contribution, 
we have considered the $\gamma d \to \vec{n} p$ 
process with the pionless EFT employing the di-baryon fields.
We have calculated the PV neutron spin asymmetry $P_{z'}$ 
as a function of the photon energy up to 30~MeV.
Our results could be useful to fix
the PV $dNN$ LECs appearing in the pionless PV Lagrangian.
\begin{figure*}
\centering
  \includegraphics[width=0.32\textwidth]{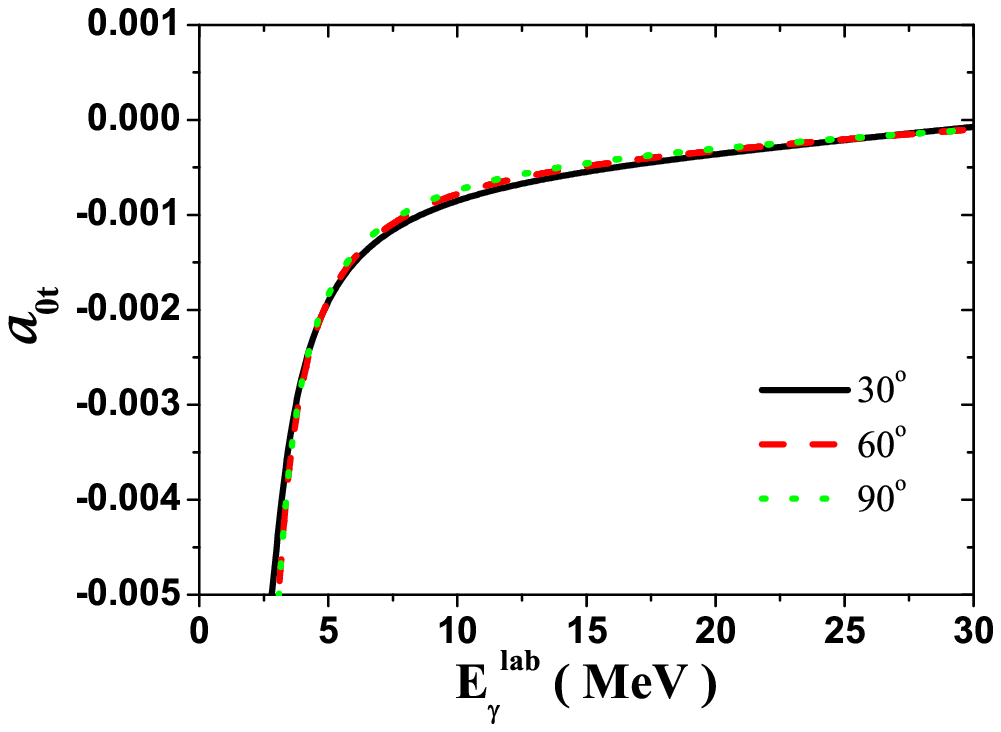}
  \includegraphics[width=0.32\textwidth]{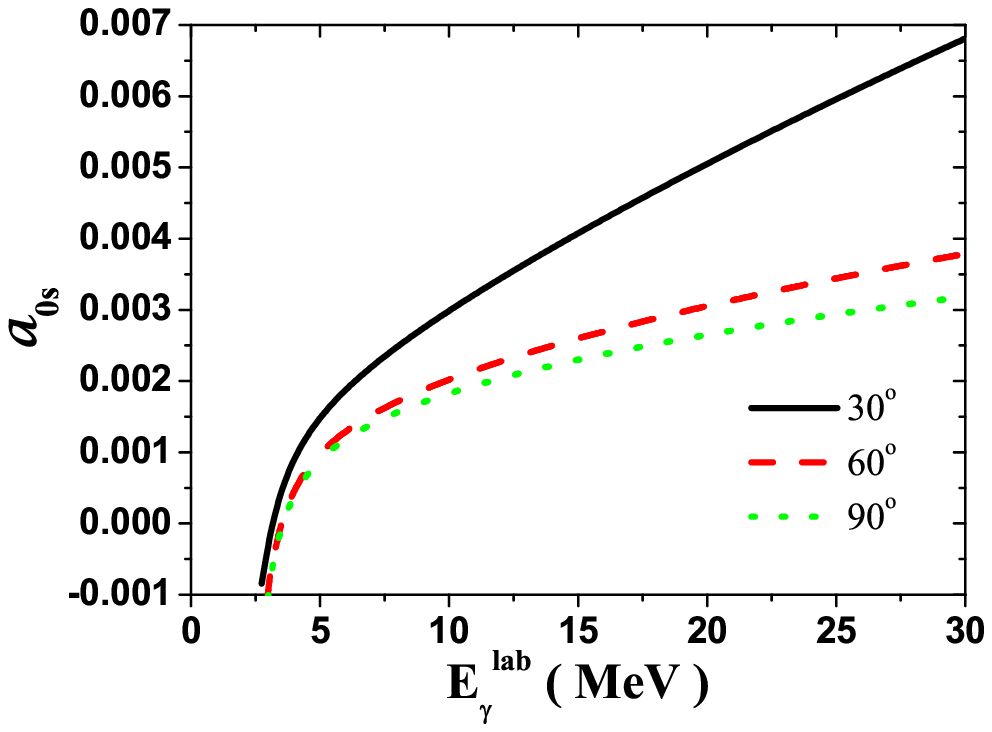}
  \includegraphics[width=0.32\textwidth]{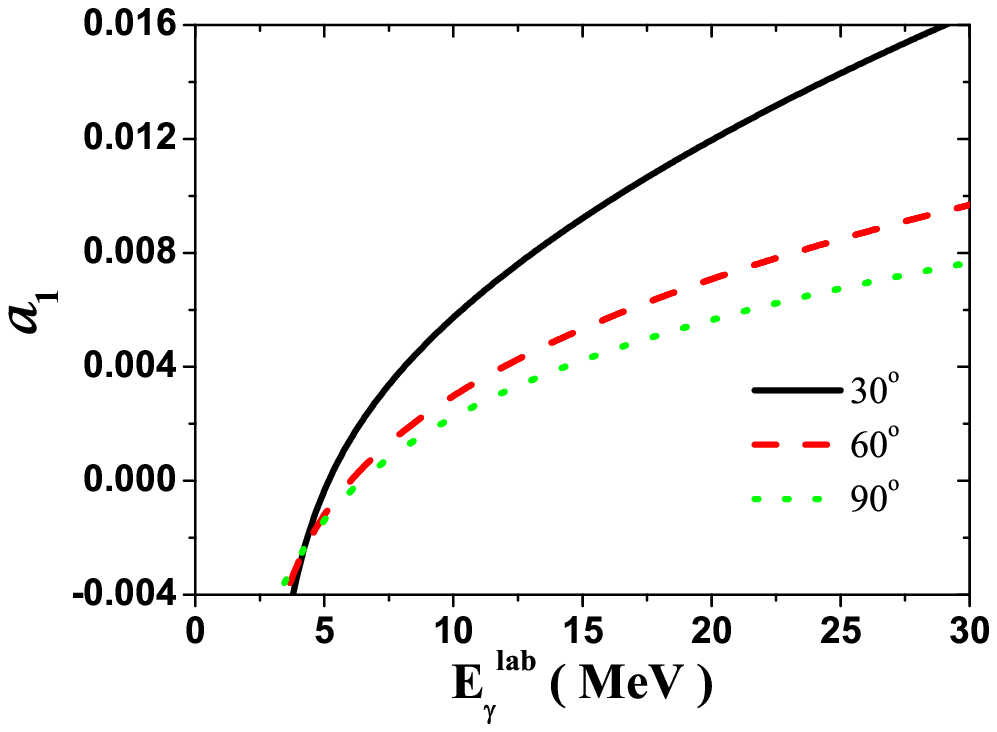}
\caption{From left $a_{0t}$, $a_{0s}$ and $a_1$ plotted 
as a function of $E^{lab}_{\gamma}$ at $\theta=30^\circ$, 
60$^\circ$, 90$^\circ$.}
\label{fig:1}       
\end{figure*}
\begin{acknowledgements}
 J.W. Shin was supported by National Nuclear R \& D Program 
(No. 2011-0006347) and S.W. Hong was supported in part by the 
WCU program (R31-2008-10029) through the National Research 
Foundation of Korea (NRF) funded by the Ministry of Education,
Science and Technology.

\end{acknowledgements}

\end{document}